\documentclass[%
 reprint,
superscriptaddress,
 amsmath,amssymb,
 aps,
prl,
longbibliography
]{revtex4-2}

\usepackage{graphicx}%
\usepackage{dcolumn}%
\usepackage{multirow}
\usepackage{amssymb}
\usepackage{comment}
\usepackage{MnSymbol,wasysym}
 \usepackage{setspace}

\usepackage[dvipsnames,table,xcdraw]{xcolor}

\usepackage{hyperref}
\hypersetup{
    colorlinks=true,
    linkcolor=Blue,
    filecolor=Blue,      
    urlcolor=Blue,
    citecolor=blue,
}

\definecolor{Gray}{gray}{0.85}
\newcolumntype{g}{>{\columncolor{Gray}}r}
\newcolumntype{w}{>{\columncolor{white}}r}

\begin{document}

\title{From pure to mixed: 
Altermagnets as intrinsic symmetry-breaking indicators}

\author{Aline Ramires}
\affiliation{Laboratory for Theoretical and Computational Physics, PSI Center for Scientific Computing, Theory, and Data,  Paul Scherrer Institute, 5232 Villigen PSI, Switzerland.}

\date{\today}

\begin{abstract}
We investigate the impact of the presence of altermagnetism on physical observables in the presence of explicit or spontaneous symmetry-breaking fields.
We focus on unconventional superconductivity as a potential source of spontaneous symmetry-breaking fields and derive the symmetry considerations for the transmutation of altermagnets from pure to mixed, which can give rise to unusual phenomena. 
Based on this analysis, we put forward scenarios that could explain the apparent onset of time-reversal symmetry breaking at the superconducting critical temperature in Sr$_2$RuO$_4$, and the hidden magnetic phase and magnetic memory in 4Hb-TaS$_2$. 
\end{abstract}

\maketitle 


Altermagnets have attracted significant attention given their status as a new class of collinear-compensated magnetic phases and their potential application in spintronic devices \cite{Smejkal2022Emerging}.
Spin space groups, generalizations of magnetic space groups that ignore spin-orbit cpoupling (SOC), were identified as a useful theoretical tool for understanding the unconventional spin splitting in the electronic bands of these materials and have successfully guided the discovery of altermagnetism in weakly-correlated systems \cite{Smejkal2022Beyond}.

In real materials, however, SOC is unavoidable and physical responses such as the anomalous Hall effect (AHE) critically depend on the coupling between spin and lattice degrees of freedom.
Importantly, despite their zero nonrelativistic total magnetization, some altermagnets develop a weak relativistic magnetization, and therefore AHE,  when the  N\`{e}el vector is compatible with a Hall vector \cite{SmejkalNatRev2022}. 
Recently, altermagnets supporting an AHE were dubbed \emph{mixed}, in contrast to \emph{pure} altermagnets that do not support an AHE \cite{Fernandes2024}. 


In Fernandes et al. \cite{Fernandes2024}, a zero or nonzero anomalous Hall response has been associated with the notions of pure and mixed altermagnets, captured by the irreducible representations (irreps) of the point group of the underlying nonmagnetic crystal. 
If the altermagnetic phase transforms in the same way as some component of a Hall vector (if they belong to the same irrep), these can emerge in superposition and the altermagnet is classified as mixed.  Otherwise, it is a pure altermagnetic phase. 
Only the former allows for an AHE. 

\begin{figure}[t]
\begin{center}
\includegraphics[width=0.4\textwidth]{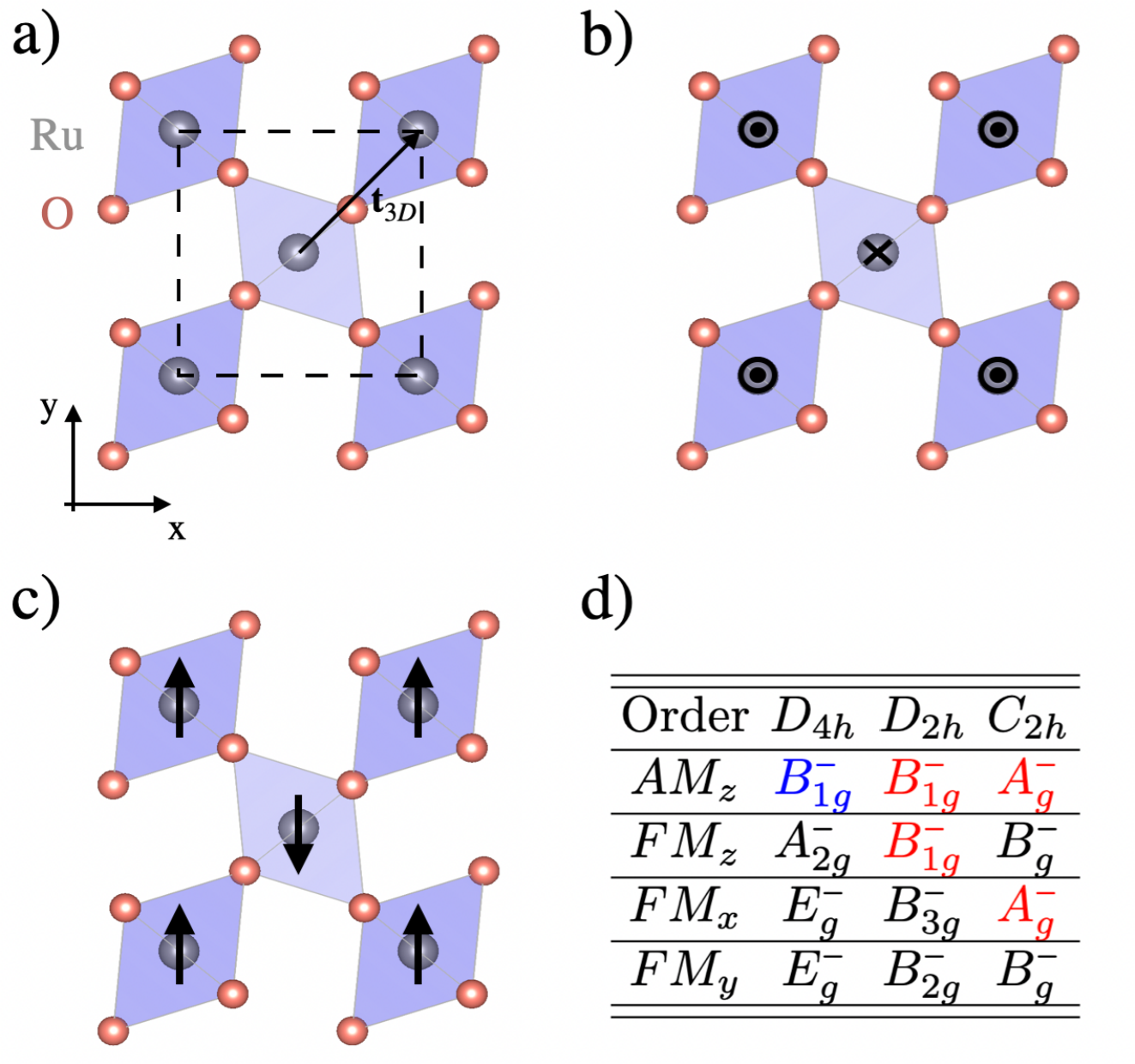}
\vspace{-0.5cm}
\end{center}
  \caption{a) Crystal structure of RuO$_2$ projected into the $xy$-plane. The oxygen octahedra are highlighted in purple (light/central and dark/outer purple octahedra lie on distinct layers). The dashed line delimits the unit cell. b) and c) Altermagnetic orders with spins along the $z$-, $y$-directions, respectively. d) Summary of irreps associated with $AM_z$ and $FM_i$ ($i=x,y,z$) orders for different point group symmetries highlighting pure (blue) and mixed (red) altermagnets.}
      \label{Fig:RuO2}
\end{figure}

RuO$_2$ is the quintessential example of an altermagnet, with crystallographic space group $P4_2/mnm$ ($\# 136$).
This material has a natural sublattice structure endowed by its nonsymmorphic character; see Fig. \ref{Fig:RuO2} (a). 
As the sublattices are not related by inversion symmetry, this crystal structure allows for a homogeneous magnetic order with antiparallel spins in each sublattice, characterizing it as an altermagnet. 
Factoring out translations, the group is isomorphic to $D_{4h}$, with the caveat that some transformations exchange sublattice labels. 

The symmetry group of RuO$_2$ is generated by $\bar{C}_{4z}$ a four-fold rotation along the $z$-axis followed by a fractional lattice translation $\mathbf{t}_{3D} = (1/2,1/2,1/2)$ (indicated by the bar); $\bar{C}_{2x}$ and $C_{2d}$, two-fold rotations along the $x$- and $d$($x=y$)-axes, respectively, the former followed by a translation by $\mathbf{t}_{3D}$; and inversion $i$. 
A collinear-compensated magnetic order with spins along the $z$-axis, labeleld as $AM_z$,  breaks $\bar{C}_{4z}$  and $C_{2d/\bar{d}}$ but preserves $\bar{C}_{2x/y}$, associating this altermagnetic order with $B_{1g}^-$ (the $-$ superscript indicates that the irrep is odd w.r.t. time-reversal symmetry), see Fig. \ref{Fig:RuO2} (b). 
Analogously, collinear-compensated magnetic order with spins along the $y$-axis, $AM_y$, is related to $AM_x$ by $\bar{C}_{4z}$ symmetry, associating these two orders with a basis for the $E_g^-$ irrep, see Fig. \ref{Fig:RuO2} (c). 
In $D_{4h}$, the magnetic field component $h_z$ transforms as $A_{2g}^-$ and $\{h_x, h_y\}$ transform as $E_g^-$, characterizing $AM_z$ as a pure altermagnet and $AM_p$ (with $p$ denoting some in-plane direction) as a mixed altermagnet.

If RuO$_2$ hosts the pure altermagnet order $AM_z$, the reorientation of the N\'eel vector away from the $z$-axis, achieved by an external magnetic field, can transform this altermagnet from pure into mixed \cite{Smejkal2020, Feng2022}. 
Note that the explicit reduction of lattice symmetries could also promote a transformation from pure to mixed altermagnet. 
In this example, reducing the underlying lattice symmetry $D_{4h} \rightarrow D_{2h}$, breaking $\bar{C}_{4z}$ and $\bar{C}_{2x/y}$, leads to the irrep descent $\{A_{2g}^-, B_{1g}^- \}\rightarrow B_{1g}^-$ and $E_{g}^- \rightarrow B_{2g}^- \oplus B_{3g}^-$, indicating that the $AM_z$ order is not a pure AM anymore, as it can acquire a finite $FM_z$ component (it is actually a ferrimagnet, as all symmetries connecting the sublattices and imposing the spin compensation are broken). 
On the other hand, the reduction of lattice symmetries following $D_{4h} \rightarrow C_{2h}$, breaking $\bar{C}_{4z}$ and ${C}_{2d/\bar{d}}$, leads to the irrep descent $A_{2g}^-\rightarrow B_{g}^-$, $B_{1g}^-\rightarrow A_{g}^-$ and $E_{g}^- \rightarrow A_{g}^- \oplus B_{g}^-$, indicating that the $AM_z$ order is a mixed altermagnet, as it is compatible with a finite $FM$ component with spins along the (in-plane) axis of the remaining  $C_2$ symmetry. Figure \ref{Fig:RuO2} (d) summarizes this discussion.

We now apply a similar analysis to the surface of Sr$_2$RuO$_4$.  
Sr$_2$RuO$_4$ has been intensively studied and characterized by various probes over the past decades, given its potential to host chiral superconductivity \cite{Maeno2024}. 
One of the key experimental indicators of such an exotic superconducting state comes from the apparent time-reversal symmetry breaking (TRSB) at the superconducting transition temperature, inferred by polar Kerr effect (PKE) \cite{Xia2006} and muon spin resonance ($\mu$SR) experiments \cite{Luke1998, Grinenko2021, Grinenko2023}. 
Curiously, recent low-energy $\mu$SR (LE$\mu$SR) experiments suggested that TRSB is already established at the surface of Sr$_2$RuO$_4$ at temperatures around $50$K, much higher than the superconducting critical temperature of about $ 1.5$K \cite{Fittipaldi2021}.
Inspired by these results, we lay out a thought experiment. We consider the possibility of altermagnetism on the surface of Sr$_2$RuO$_4$, characterize it as pure or mixed,  and study the consequences of its presence to surface responses with the onset of unconventional superconductivity.

\begin{figure}[t]
\begin{center}
\includegraphics[width=0.4\textwidth]{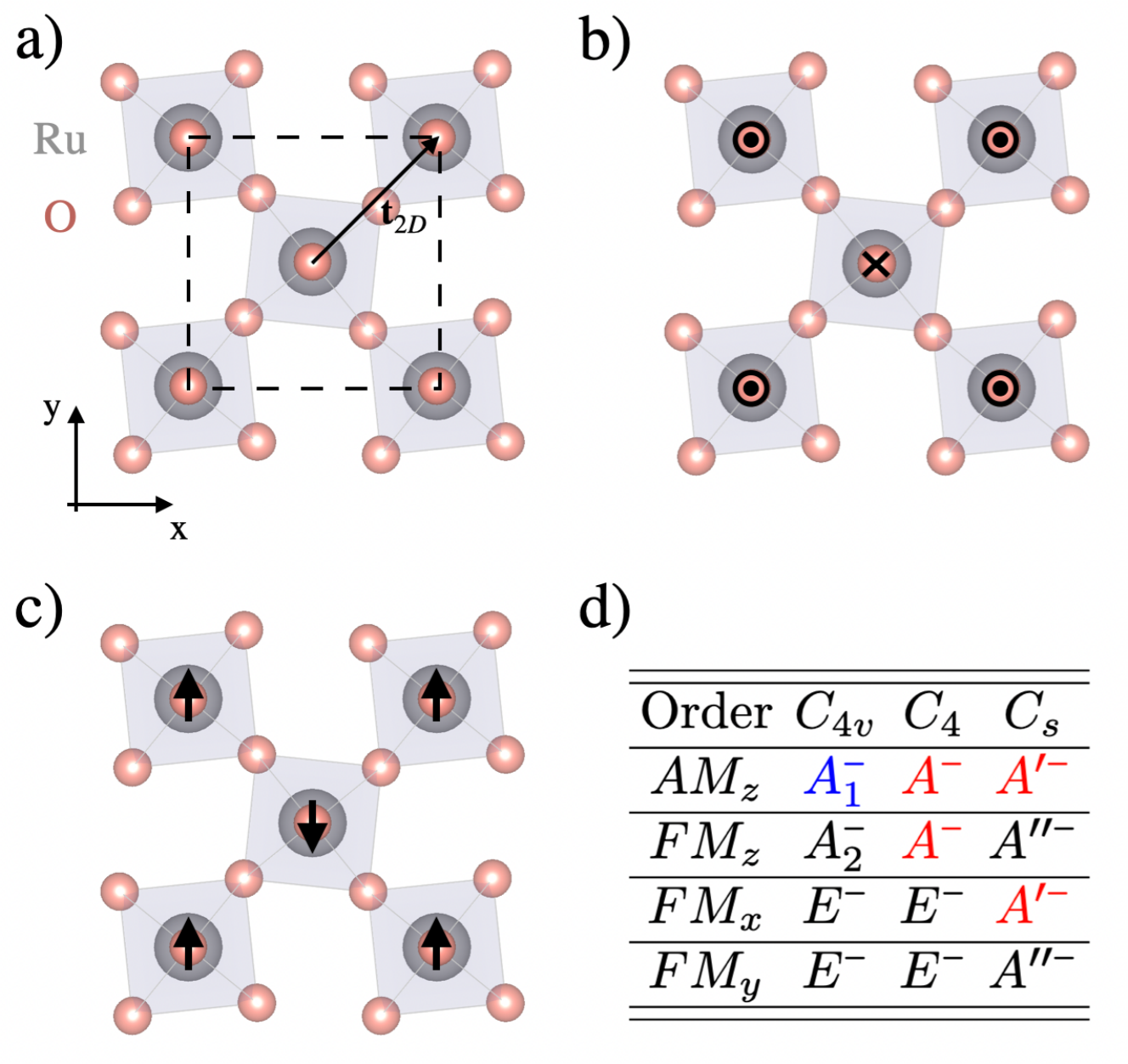}
\end{center}
\vspace{-0.5cm}
  \caption{a) Crystal structure of the (001) surface of Sr$_2$RuO$_4$.  b) and c) Altermagnetic orders with spins along the $z$- and $y$-axis, respectively. d) Summary of irreps associated with $AM_z$ and $FM_i$ ($i=x,y,z$) orders for different point group symmetries.}
      \label{Fig:SRO}
\end{figure}

Bulk Sr$_2$RuO$_4$ has $I4/mmm$  ($\# 139$) space group symmetry, with a body-centered tetragonal unit cell containing aligned oxygen tetrahedra surrounding each Ru atom. 
The surface, however, is not a simple projection of the bulk, as the oxygen tetrahedra are known to be rotated at the surface \cite{Matzdorf2000, Matzdorf2002}, see Fig. \ref{Fig:SRO} (a). 
With this distortion, the plane space group is $p4gm$ ($\#12$), generated by $C_{4z}$, a four-fold rotational symmetry along the $z$-axis (out of the plane); $\bar{M}_x$, a mirror symmetry with normal along the $x$-axis; and $\bar{M}_d$, a glide plane with normal along the $d$-axis, the last two operations accompanied by a shift by half a reciprocal lattice vector $\mathbf{t}_{2D} = (1/2,1/2)$ (indicated by the bar).
The nonsymmorphic nature of this symmetry group guarantees the presence of a sublattice structure within which altermagnetic order can emerge. 

The point group at the surface is isomorphic to $C_{4v}$. $AM_z$ is associated with the  $A_1^-$ irrep, see Fig. \ref{Fig:SRO} b), while in-plane altermagnetic orders, such as $AM_y$, are associated with $E^-$, see Fig. \ref{Fig:SRO} c).
Hall vectors along the $z$-axis or in-plane are associated with $A_2^-$ and $E^-$ irreps, respectively, characterizing $AM_z$ as a pure altermagnet and colinear compensated magnetic orders with in-plane spins as mixed altermagnets.  
Reducing spatial symmetries such that $C_{4v} \rightarrow C_4$, the irrep descent follows $\{A_1^-, A_2^-\} \rightarrow A^-$, $\{B_1^-, B_2^-\} \rightarrow B^-$ and $E^-\rightarrow E^-$, such that $AM_z$ is not a pure altermagnet anymore (it is a ferrimagnet). 
The reduction of spatial symmetries such that only one mirror symmetry is preserved, $C_{4v} \rightarrow C_{s}$, choosing the mirror with normal along the $x$-axis, the irreps descent following $\{A_1^-, B_1^-\} \rightarrow A'^-$, $\{A_2^-, B_2^-\} \rightarrow A''^-$ and $E^-\rightarrow A'^- \oplus A''^-$. 
This indicates that under this symmetry reduction, $AM_z$ is a mixed altermagnet. The discussion above is summarized in Fig. \ref{Fig:SRO} d).

Selective spatial symmetry breaking implemented by the application of uniaxial strain has already been experimentally achieved and theoretically investigated in the context of Sr$_2$RuO$_4$ \cite{Hicks2014, Sunko2019, Luo2019, Barber2019, Jerzembeck2022, Jerzembeck2023, Grinenko2023, Ramires2017, Romer2020, Acharya2021, Roising2022,  Beck2022, Yuan2023}. 
The application of strain in the $B_{1g}$ or $B_{2g}$ channels breaks the four-fold rotational symmetry and a set of glide planes but does not allow for the development of $FM_i$ contributions to the pure $AM_z$ order parameter. 
Nonetheless, applying uniaxial strain in both channels concomitantly would introduce enough symmetry breaking to allow for a finite Hall vector along the $z$-direction. 
For the emergence of a Hall vector with in-plane components, the reminiscent two-fold rotational symmetry along the $z$-axis must be broken, which could be achieved by the application of an in-plane gate or the presence of structural defects, such as terraces, on the surface. 
Going beyond external symmetry-breaking fields, the onset of secondary orders can also break symmetries and allow the transmutation of a pure altermagnet into a mixed altermagnet.

If the secondary phase is superconducting, it can couple to magnetic order parameters only through gauge-invariant bilinears. If the superconducting order parameter is associated with a one-dimensional irrep, any bilinear belongs to the trivial representation, breaking no spatial symmetries, and, therefore, cannot transmute an altermagnet from pure into mixed. 
Nevertheless, an order parameter with two components allows for the construction of bilinears which break lattice symmetries. 
More specifically, a two-component superconducting order parameter, here labeled as $(\Delta_x, \Delta_y)$, allowing for a non-zero gauge-invariant bilinears $|\Delta_x|^2-|\Delta_y|^2$ and $\Delta_x\Delta_y^* +  \Delta_y^*\Delta_x$, could allow for the onset of a finite $FM_i$ on the surface with the onset of superconductivity.
At the free energy level, this would be accounted for by the presence of terms $\sim AM_z. FM_i. |\Psi_B|^2 $, where $|\Psi_B|^2$ is the appropriate superconductig bilinear and $i=x,y,z$. 

For the example at hand,  $FM_z$ would be induced if $|\Psi_B|^2$  belongs to the $A_2$ irrep. This requires $\Delta_x\Delta_y^* +  \Delta_y^*\Delta_x$ to be non-zero, and therefore is associated with a nematic order parameter. Note that this bilinear naively transforms as $B_{2}$, but if one considers also higher angular momentum components of the gap, this product decomposes in to a sum of $B_{2}$ and $A_{2}$ terms. The induction of $FM_{x,y}$ by a superconducting bilinear would be possible if this breaks all spatial symmetries but one mirror plane with in-plane normal, or a two-fold rotation with in-plane axis. A complete table highlighting more details of the symmetry descent is given in the supplemental material (SM).
Importantly, note that within this mechanism the superconducting state does not need to intrinsically break TRS for the onset of $FM_i$ and the concomitant onset of a  nonzero PKE response.

The discussion above suggests the following scenario: the surface of Sr$_2$RuO$_4$ hosts a pure altermagnet, $FM_z$, above the superconducting critical temperature (corroborated by the LE$\mu$SR experiments \cite{Fittipaldi2021} and the zero PKE \cite{Xia2006}); with the onset of superconductivity in the bulk spatial symmetries are broken in such as way that the altermagnet on the surface is transformed from pure into mixed, allowing for a finite PKE. What remains to be reconciled in this scenario are the reports from bulk $\mu$SR experiments \cite{Luke1998, Grinenko2021, Grinenko2023}. These experiments use muons of enough energy which are believed to penetrate into the sample to depths of about 0.1mm  \cite{Luke1998, Grinenko2021}. Note, however, that the local fields perceived by the muons could be actually originated from the surface. Surface roughness allow surface stray fields to penetrate the bulk up to length scales comparable to the characteristic roughness length scale \cite{Demokritov1994, Tsymbal1994}. 
STM studies have reported extended terraces with characteristic separation of about 10$\mu$m \cite{Matzdorf2000, Pennec2008}, and Kerr effect measurements have suggested the presence of superconducting domains of about 50-100$\mu$m  \cite{Xia2006, Kallin2016}. The change in the inhomogeneity length scales as the system goes from the normal to the superconducting state could be the reason behind the apparent onset of TRSB in the bulk reported by $\mu$SR.

Note that the scenario above is also in agreement with $\mu$SR experiments under in-plane uniaxial strain. If the strain breaks $C_{4z}$ symmetry, the superconducting order parameter is now associated with a one-dimensional irrep and cannot break spatial symmetries just below the superconducting critical temperature, and the surface $AM_z$ order remains pure. At lower temperatures, with the onset of the second superconducting component, the  superconducting order  parameter can now break spatial symmetries and domains can be formed. These two features would then imply that at a lower temperature a Hall vector detectable by PKE could develop and that domains with a characteristic length scale comparable to the one in absence of strain would form and allow stray fields to penetrate deeper into the sample and be perceived by bulk muons.

The LE$\mu$SR experiments in Sr$_2$RuO$_4$ inspiring the discussion above are subtle as the magnetic signal reported is very small \cite{Fittipaldi2021}. 
Consistent with these results are STM experiments reporting an inequivalent local density of states on two crystallographically equivalent Sr sites \cite{Pennec2008, Marques2021}. 
Ascribing the TRSB to surface altermagnetism would also be in agreement with the absence of bulk thermodynamical signatures of a secondary phase transition associated with the onset of TRSB in strained Sr$_2$RuO$_4$  \cite{Li2022, Jerzembeck2024}. 
From a theoretical perspective, recent first-principles calculations have shown that the rotation of the oxygen octahedra in Sr$_2$RuO$_4$ leads to the stabilization of altermagnetism \cite{Autieri2025}. Similar results were found for Ca$_2$RuO$_4$, with tilted and rotated octahedra in the bulk \cite{Cuono2024, Gonzlez2025}. Finally, La$_2$CuO$_4$, which bulk hosts rotated octahedra in the bulk, has also been suggested to be an altermagnet based on first-principles calculations \cite{Lane2018,Smejkal2022Emerging}.


This mechanism could also be behind the unusual interplay between magnetism and superconductivity in 4Hb-TaS$_2$. Scanning superconducting quantum interference device (SQUID) microscopy experiments have reported a spontaneous vortex phase in this material, which suggests a hidden magnetic order above the superconducting transition temperature \cite{Persky2022}. Nevertheless, the magnetic order has not been detected using conventional techniques and seems to be incompatible with ferromagnetism \cite{Persky2022}. Muon spin resonance experiments have indicated the onset of TRSB in this system only at the superconducting transition, suggesting that this material hosts a chiral superconducting state \cite{Ribak2020}. Furthermore, the reported two-fold symmetry of the superconducting critical field suggests a nematic superconducting order parameter \cite{Silber2024}. Inspired by these results, we investigate whether the mechanism proposed here could also provide a resolution to this complex phenomenology.

\begin{figure}[t]
\begin{center}
\includegraphics[width=0.425\textwidth]{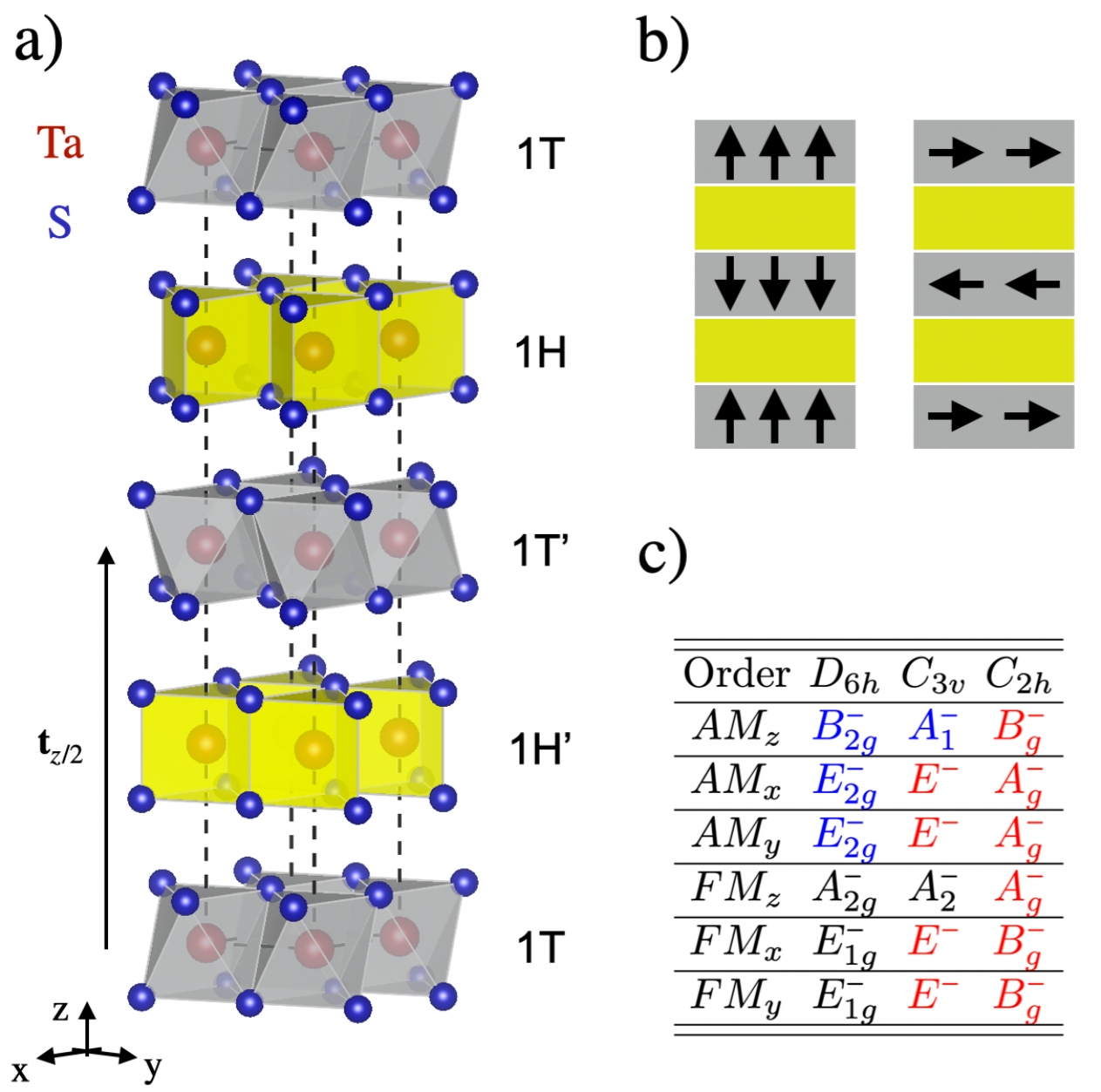}
\end{center}
\vspace{-0.5cm}
  \caption{a) 4Hb-TaS$_2$ crystal structure highlighting the 1T and 1H layers with gray and yellow polyhedra, respectively.   b) Schematic depiction on a simplified structure of $AM_z$ (left) and $AM_p$ (right) orders.
  c) Summary of irreps associated with $AM_i$ and $FM_i$ ($i= x,y,z$) orders for diﬀerent point group symmetries.}
      \label{Fig:TaS2}
\end{figure}

4Hb-TaS$_2$ is a hexagonal system with space group $P6_3/mmc$ ($\#$ 194), with a unit cell consisting of alternating layers of 1H- and 1T-TaS$_2$, as depicted in Fig. \ref{Fig:TaS2} a). The group is generated by $\bar{C}_{6z}$, a six-fold rotational symmetry along the $z$-axis followed by translation by $\boldsymbol{t}_{z/2} = (0,0,1/2)$ (denoted by the bar); $C_{2x}$ and $\bar{C}_{2y}$, two-fold rotations along hte $x$- and $y$-axes, respectively, with the latter followed by a shift by $\boldsymbol{t}_{z/2}$; and inversion $i$. This space group is also nonsymmorphic, featuring a screw symmetry along the $z$-axis, which endows the structure with a sub-layer structure, analogous to the sublattice structures discussed above.

The corresponding point group is isomorphic to $D_{6h}$,  with the caveat that some of the operations exchange layers labelled as $T$ and $T'$ (we focus the discussion on the $1T$ layers as these are the layers where magnetism is expected to emerge). Altermagnetism in this system could emerge in the form of compensated ferromagnetic layers, as depicted in Fig. \ref{Fig:TaS2} b). Altermagnetic order with moments aligned along the $z$-axis, $AM_z$, is associated with the $B_{2g}^-$ irrep, while altermagnetic orders with in-plane moments are associated with the $E_{2g}^-$ irrep. Hall vectors along the $z$-axis or in-plane are associated with the $A_{2g}^-$ and $E_{1g}^-$ irreps, respectively. This qualifies all altermagnetic orders in this system as pure altermagnets. 

The reduction of spatial symmetries could transmute these altermagnetic phases from pure to mixed. 
Reducing the spatial symmetry such that $D_{6h} \rightarrow C_{3v}$, the irrep descent follows $\{A_{1g}^-, B_{2g}^-\}\rightarrow A_1^-$,  $\{A_{2g}^-, B_{1g}^-\}\rightarrow A_2^-$, and $\{E_{1g}^-,E_{2g}^-\}\rightarrow E_g^-$, indicating that $AM_z$ remains pure, but $AM_p$ becomes a mixed altermagnet.
Breaking both three- and six-fold rotational symemtries such that $D_{6h}\rightarrow C_{2h}$, the irrep descent follows $\{A_{1g}^-, A_{2g}^-, E_{2g}^-\}\rightarrow A_g^-$ and  $\{B_{1g}^-, B_{2g}^-, E_{2g}^-\}\rightarrow B_g^-$, such that all altermagnets are transmuted from pure to mixed. A summary of these cases is given in Fig. \ref{Fig:TaS2} c) and a complete symmetry descent table can be found in the SM.

Going back to the tantalizing experimental results in this material in the light of this symmetry analysis, one scenario that accounts for the apparent onset of TRSB at the superconducting transition and the magnetic memory effect is the following: 4Hb-TaS$_2$ hosts a pure altermagnetic order above the superconducting transition temperature which cannot be captured by SQUID microscopy \cite{Persky2022}. 
If the superconducting order parameter has two components, it can lead to spatial symmetry breaking captured by gauge-invariant bilinears as discussed above for the case of Sr$_2$RuO$_4$. 
This fact is supported by the two-fold anisotropic superconducting critical field suggesting a nematic superconducting state \cite{Silber2024}. 
Breaking the three- and six-fold rotational symmetries in this system leads to the transmutation of all altermagnetic states from pure to mixed. The magnetic state can then acquire a finite ferromagnetic component, what is seen as an extra internal field by the muons onsetting at the superconducting transition \cite{Ribak2020}. 
Again, in this scenario the TRSB is not intrinsic to the superconducting state, but hidden in the underlying pure altermagetic state. 
The TRSB becomes evident only once spatial symmetries are broken and the altermagnets transmute from pure to mixed.

Assuming the magnetic state acquires a ferromagnetic component below the superconducting critical temperature, one can expect that this ferromagnetic component will behave as a standard ferromagnet, and manifest magnetic hysteresis, as reported experimentally \cite{Persky2022}. 
Furthermore, the magnetic memory effect could be explained by the interplay between the energies associated with vortex formation, and vortex-ferromagnetic domain interaction, which would require a more detailed study.


In summary, we investigated the impact of the presence of altermagnetism on physical obervables in the presence of extra symmetry-breaking fields introduced by the onset of superconducting order parameters.
In the context of Sr$_2$RuO$_4$, we hope the picture for the origin of the TRSB proposed here will motivate further theoretical studies and experimental characterization of the potential magnetic orders emerging at its surface. This scenario could be tested by passivating the surface to avoid the octahedral rotation, and, therefore, the emergence of altermagnetism and the subsequent apparent onset of TRSB with superconductivity.
For the case of 4Hb-TaS$_2$, we also propose a plausible scenario for the hidden magnetic order and trainable magnetic memory, which does not rely on exotic magnetic or superconducting states \cite{Liu2024, Konig2024, Lin2024}. This scenario could be checked by the application of strain, explicitly breaking three-fold symmetry, what should lead to the emergence of a detectable FM component to this altermagnet.
The two case scenarios discussed here highlight the generality of the altermagnet transmutation mechanism for the apparent onset of TRSB and can be potentially relevant to many other materials.

\acknowledgements The author acknowledges support from the Swiss National Science Foundation through Ambizione Grant No. 186043. The author thanks Andras Szabo, Andriy Smolyanyuk, Hubertus Luetkens, Karsten Held, Manfred Sigrist, and Roustem Khassanov for inspiring discussions.

\bibliography{AMSurfaceSRO_R1}{}

\end{document}